\documentclass[english,numreferences]{kluwer}
\usepackage{graphicx}
\usepackage{amssymb}

\makeatletter

\usepackage{klucite}

\usepackage{babel}
\makeatother
\begin{document}

\newcommand{\commute}[2]{\left[#1,#2\right]}

\newcommand{\bra}[1]{\left\langle #1\right|}

\newcommand{\ket}[1]{\left|#1\right\rangle }

\newcommand{\anticommute}[2]{\left\{  #1,#2\right\}  }
\begin{article}

\begin{opening}

\title{Non-Markovian dynamics of a localized electron spin due to the hyperfine
interaction}

\author{W. A.  \surname{Coish} and Daniel  \surname{Loss}.}

\institute{University of Basel, Klingelbergstrasse 82, CH-4056 Basel, Switzerland}

\runningauthor{W. A. Coish and Daniel Loss}

\begin{abstract}
We review our theoretical work on the dynamics of a localized electron
spin interacting with an environment of nuclear spins. Our perturbative
calculation is valid for arbitrary polarization $p$ of the nuclear
spin system and arbitrary nuclear spin $I$ in a sufficiently large
magnetic field. In general, the electron spin shows rich dynamics,
described by a sum of contributions with exponential decay, nonexponential
decay, and undamped oscillations. We have found an abrupt crossover
in the long-time spin dynamics at a critical shape and dimensionality
of the electron envelope wave function. We conclude with a discussion
of our proposed scheme to measure the relevant dynamics using a standard
spin-echo technique.
\end{abstract}


\end{opening}

\section{Introduction}

Experiments with trapped ions\cite{kielpinski:2001a} and nuclear
magnetic resonance (NMR)\cite{vandersypen:2001a} have proven that
the basic elements of a quantum computer can be realized in practice.
To progress beyond proof-of-principle experiments, the next generation
of quantum information processors must overcome significant obstacles
regarding scalability and decoherence. The scalability issue is largely
solved by proposals for a solid-state implementation for quantum computing,
where established fabrication techniques can be used to multiply the
qubits and interface them with existing electronic devices. Due to
their relative isolation from the surrounding environment, single
electron spins in semiconductor quantum dots are expected to be exceptionally
robust against decoherence.\cite{loss:1998a} Indeed, longitudinal
relaxation times in these systems have been measured to be $T_{1}=0.85\,\mathrm{ms}$
in a magnetic field of $8\,\mathrm{T}$\cite{elzerman:2004a} and
in GaAs quantum wells, the transverse dephasing time $T_{2}^{*}$
for an ensemble of electron spins (which typically provides a lower
bound for the intrinsic decoherence time $T_{2}$ of an isolated spin)
has been measured to be in excess of $100\,\mathrm{ns}$.\cite{kikkawa:1998a} 

Possible sources of decoherence for a single electron spin confined
to a quantum dot are spin-orbit coupling and the contact hyperfine
interaction with the surrounding nuclear spins.\cite{burkard:1999a}
The relaxation rate due to spin-orbit coupling $1/T_{1}$ is suppressed
for localized electrons at low temperatures\cite{khaetskii:2000a}
and recent work has shown that $T_{2}$, due to spin-orbit coupling,
can be as long as $T_{1}$ under realistic conditions.\cite{golovach:2004a}
However, since spin-carrying isotopes are common in the semiconductor
industry, the contact hyperfine interaction (in contrast to the spin-orbit
interaction) is likely an unavoidable source of decoherence, which
does not vanish with decreasing temperature or carefully chosen quantum
dot geometry.\cite{schliemann:2003a} 

In the last few years, a great deal of effort has been focused on
a theoretical description of interesting effects arising from the
contact hyperfine interaction for a localized electron. The predicted
effects include a dramatic variation of $T_{1}$ with gate voltage
in a quantum dot near the Coulomb blockade peaks or valleys,\cite{lyanda-geller:2002a}
all-optical polarization of the nuclear spins,\cite{imamoglu:2003a}
use of the nuclear spin system as a quantum memory,\cite{taylor:2003a,taylor:2003b}
and several studies of spin dynamics.\cite{khaetskii:2002a,erlingsson:2001a,erlingsson:2002a,semenov:2004a,desousa:2003a,yuzbashyan:2004a,desousa:2004a,shenvi:2004a}
Here, our system of interest is an electron confined to a single GaAs
quantum dot, but this work applies quite generally to other systems,
such as electrons trapped at shallow donor impurities in Si:P.\cite{schliemann:2003a}

An exact solution for the electron spin dynamics has been found in
the special case of a \emph{fully} polarized initial state of the
nuclear spin system.\cite{khaetskii:2002a,khaetskii:2003a} This solution
shows that the electron spin only decays by a fraction $\propto\frac{1}{N}$
of its initial value, where $N$ is the number of nuclear spins within
the extent of the electron wave function. The decaying fraction was
shown to have a nonexponential tail for long times, which suggests
non-Markovian (history dependent) behavior. For an initial nuclear
spin configuration that is not fully polarized, no exact solution
is available and standard time-dependent perturbation theory fails.\cite{khaetskii:2002a}
Subsequent exact diagonalization studies on small spin systems\cite{schliemann:2002a}
have shown that the electron spin dynamics are highly dependent on
the type of initial nuclear spin configuration. The unusual (nonexponential)
form of decay, and the fraction of the electron spin that undergoes
decay may be of interest in quantum error correction (QEC) since QEC
schemes typically assume exponential decay to zero. 

In the following section we describe a systematic perturbative theory
of electron spin dynamics under the action of the Fermi contact hyperfine
interaction. Further details of this work can be found in reference
\cite{coish:2004a}.

\section{Model and perturbative expansion}

We investigate electron spin dynamics at times shorter than the nuclear
dipole-dipole correlation time $\tau_{\mathrm{dd}}$ ($\tau_{\mathrm{dd}}\approx10^{-4}\,\mathrm{s}$
in GaAs is given directly by the inverse width of the nuclear magnetic
resonance (NMR) line\cite{paget:1977a}). At these time scales, the
relevant Hamiltonian for a description of the electron and nuclear
spin dynamics is that for the Fermi contact hyperfine interaction:
\begin{eqnarray}
H & = & H_{0}+V_{\mathrm{ff}},\label{eq:HFHamiltonian}\\
H_{0} & = & \left(b+h_{z}\right)S_{z};\,\,\, V_{\mathrm{ff}}=\frac{1}{2}\left(S_{+}h_{-}+S_{-}h_{+}\right),\end{eqnarray}
where $\mathbf{S}$ is the electron spin operator, $b$ gives the
electron Zeeman splitting and $\mathbf{h}=\sum_{k}A_{k}\mathbf{I}_{k}$
is the quantum nuclear field. $S_{\pm}=S_{x}\pm iS_{y}$ and $h_{\pm}=h_{x}\pm ih_{y}$
are defined in the usual way. The hyperfine coupling constants are
$A_{k}=Av_{0}|\psi(\mathbf{r}_{k})|^{2}$ where $v_{0}$ is the volume
of a crystal unit cell containing one nuclear spin, $\psi(\mathbf{r})$
is the electron envelope wave function, and $A$ is the strength of
the hyperfine coupling. While the total number of nuclear spins in
the system, $N_{\mathrm{tot}}$, may be very large, there are fewer
spins ($N<N_{\mathrm{tot}}$) within the extent of the electron wave
function. In GaAs, all naturally occurring isotopes carry spin $I=\frac{3}{2}$.
In bulk GaAs, $A$ has been estimated\cite{paget:1977a} to be $A=90\,\mu eV$.
This estimate is based on an average over the hyperfine coupling constants
for the three nuclear isotopes $\mathrm{^{69}Ga}$, $\mathrm{^{71}Ga}$,
and $\mathrm{^{75}As}$, weighted by their relative abundances. Natural
silicon contains 4.7\% $^{29}\mathrm{Si}$, which carries $I=\frac{1}{2}$,
and 95\% $^{28}\mathrm{Si}$, with $I=0$. An electron bound to a
phosphorus donor impurity in natural Si:P interacts with $N\approx10^{2}$
surrounding $\mathrm{^{29}Si}$ nuclear spins, in which case the hyperfine
coupling constant is on the order of $A\approx0.1\,\mu eV$.\cite{schliemann:2003a} 

For large magnetic fields $b$, the flip-flop processes due to $V_{\mathrm{ff}}$
are suppressed by the electron Zeeman splitting, and it is reasonable
to take $H\approx H_{0}$ (zeroth order in $V_{\mathrm{ff}}$). The
zeroth-order problem is algebraically simple, and leads to some interesting
conclusions regarding the initial conditions. In this limit the longitudinal
spin is time-independent, since $\commute{S_{z}}{H_{0}}=0$, but the
transverse spin can undergo nontrivial evolution. Assuming uniform
hyperfine coupling constants ($A_{k}=A/N$) and nuclear spin $I=1/2$
we evaluate the transverse electron spin dynamics for a nuclear spin
bath of polarization $p$ along the $z$-axis and two different nuclear
spin initial states. When the $A_{k}$ are uniform, the transverse
spin exhibits periodic dynamics. However, at times much less than
the period, given by the inverse level spacing ($t\ll N/A$, setting
$\hbar=1$), we find \begin{eqnarray}
\left\langle S_{+}\right\rangle _{t}^{\mathrm{unprep.}} & = & \left\langle S_{+}\right\rangle _{0}e^{i\omega_{n}t}e^{-t^{2}/2t_{c}^{2}},\,\,\,\,\, t_{c}=\frac{2}{A}\sqrt{\frac{N}{1-p^{2}}},\label{eq:SplusUnPrep}\\
\left\langle S_{+}\right\rangle _{t}^{\mathrm{prep.}} & = & \left\langle S_{+}\right\rangle _{0}e^{i\omega_{n}t},\,\,\,\,\,\omega_{n}=b+pA/2.\label{eq:SplusPrep}\end{eqnarray}
 In (\ref{eq:SplusUnPrep}), the nuclear spin system is {}``unprepared''.
This state corresponds to either a translationally invariant direct-product
state of the form $\prod_{k}\left(\sqrt{f_{\uparrow}}\ket{\uparrow}_{k}+\sqrt{1-f_{\uparrow}}\ket{\downarrow}_{k}\right)$,
where $p=2f_{\uparrow}+1$, or to a statistical mixture of product
states of the form $\ket{\uparrow\downarrow\uparrow\uparrow\cdots}$,
with average polarization $p$. In contrast, (\ref{eq:SplusPrep})
corresponds to the case when the nuclear system has been {}``prepared''
in an eigenstate $\ket{n}$ of the operator $h_{z}$: $h_{z}\ket{n}=[h_{z}]_{nn}\ket{n}$,$([h_{z}]_{nn}=pA/2)$.
It is important to note that the decay of the {}``unprepared'' state
is not irreversible. This decay can be recovered by performing a spin-echo
measurement on the electron spin, since a $\pi$-rotation of the electron
spin $S_{z}\to-S_{z}$ reverses the sign of $H_{0}$: $H_{0}\to-H_{0}$,
and results in time-reversed dynamics. When the hyperfine coupling
constants are allowed to vary in space, higher-order corrections in
$V_{\mathrm{ff}}$ will, however, result in irreversible decay. In
the following, we generalize to nonuniform $A_{k}$ and arbitrary
nuclear spin $I$ to evaluate this decay for an initial nuclear state
that is an eigenstate of the $h_{z}$-operator. 

An exact generalized master equation (GME) can be derived for the
electron spin operators beginning from the von Neumann equation for
the full density operator ($\dot{\rho}=-i\commute{H}{\rho}$).\cite{fick:1990a}
The resulting GME is expanded in terms of $V_{\mathrm{ff}}$ through
direct application of the Dyson identity. We find, quite remarkably,
that the equations of motion for the longitudinal ($\left\langle S_{z}\right\rangle _{t}$)
and transverse ($\left\langle S_{+}\right\rangle _{t}=\left\langle S_{x}\right\rangle _{t}+i\left\langle S_{y}\right\rangle _{t}$)
spin are decoupled to \emph{all orders} in $V_{\mathrm{ff}}$ and
take the form\begin{eqnarray}
\dot{\left\langle S_{z}\right\rangle }_{t} & = & N_{z}(t)-i\int_{0}^{t}dt^{\prime}\Sigma_{zz}(t-t^{\prime})\left\langle S_{z}\right\rangle _{t^{\prime}},\label{eq:SzGME}\\
\dot{\left\langle S_{+}\right\rangle }_{t} & = & i\omega_{n}\left\langle S_{+}\right\rangle _{t}-i\int_{0}^{t}dt^{\prime}\Sigma_{++}(t-t^{\prime})\left\langle S_{+}\right\rangle _{t^{\prime}}.\label{eq:SPlusGME}\end{eqnarray}
 These integro-differential equations can be converted to a pair of
algebraic equations by Laplace transformation $f(s)=\int_{0}^{\infty}dse^{-st}f(t),\,\mathrm{Re[s]>0}$.
The algebraic equations yield

\begin{equation}
S_{z}(s)=\frac{\left\langle S_{z}\right\rangle _{0}+N_{z}(s)}{s+i\Sigma_{zz}(s)};\,\,\, S_{+}(s)=\frac{\left\langle S_{+}\right\rangle _{0}}{s-i\omega_{n}+i\Sigma_{++}(s)}.\end{equation}
 The numerator term $N_{z}(s)$ and self-energy $\Sigma_{zz}(s)$
are related to the self-energy matrix elements for spin-up and spin-down
by $N_{z}(s)=-\frac{i}{2s}\left(\Sigma_{\uparrow\uparrow}(s)+\Sigma_{\uparrow\downarrow}(s)\right)$
and $\Sigma_{zz}(s)=\Sigma_{\uparrow\uparrow}(s)-\Sigma_{\uparrow\downarrow}(s)$.
All self-energy matrix elements are written in terms of an expansion
in powers of $V_{\mathrm{ff}}$: $\Sigma_{\alpha\beta}(s)=\Sigma_{\alpha\beta}^{(2)}(s)+\Sigma_{\alpha\beta}^{(4)}(s)+\cdots$,
$\alpha,\beta=(+,\uparrow,\downarrow)$. For a sufficiently large
Zeeman splitting $b$ we find that all higher-order self-energy matrix
elements are suppressed by a smallness parameter $\Delta$: \begin{equation}
\Sigma_{\alpha\beta}^{(2(k+1))}(s)\propto\Delta^{k};\,\,\,\Delta=\frac{A}{2(b+pIA)}.\end{equation}
 In Born approximation (second order in the flip-flop terms $V_{\mathrm{ff}}$),
and for an initial nuclear spin system with uniform polarization,
we find \begin{eqnarray}
\Sigma_{\uparrow\uparrow}(s) & \approx & \Sigma_{\uparrow\uparrow}^{(2)}(s)=-iNc_{+}\left[I_{+}(s-i\omega_{n})+I_{-}(s+i\omega_{n})\right],\label{eq:SigUpUpBornUniformPolarization}\\
\Sigma_{\uparrow\downarrow}(s) & \approx & \Sigma_{\uparrow\downarrow}^{(2)}(s)=iNc_{-}\left[I_{-}(s-i\omega_{n})+I_{+}(s+i\omega_{n})\right],\label{eq:SigUpDownBornUniformPolarization}\\
\Sigma_{++}(s) & \approx & \Sigma_{++}^{(2)}(s)=-iN\left[c_{-}I_{+}(s)+c_{+}I_{-}(s)\right],\label{eq:SigPlusPlusBornUniformPolarization}\\
c_{\pm} & = & I(I+1)-\frac{1}{N_{\mathrm{tot}}}\sum_{k}\left\langle I_{k}^{z}(I_{k}^{z}\pm1)\right\rangle _{0},\\
I_{\pm}(s) & = & \frac{1}{4N}\sum_{k}\frac{A_{k}^{2}}{s\mp i\frac{A_{k}}{2}}\approx\frac{d}{m}\int_{0}^{1}dx\frac{x\left|\ln x\right|^{\nu}}{s\mp ix},\,\,\,\,\,\nu=\frac{d}{m}-1.\label{eq:IPlusMinusLaplaceTrans}\end{eqnarray}
 In evaluating (\ref{eq:IPlusMinusLaplaceTrans}), we have assumed
the electron is in its orbital ground state, described by an isotropic
envelope wave function of the form $\psi(r_{k})=\psi(0)\exp\left[-\frac{1}{2}\left(\frac{r_{k}}{l_{0}}\right)^{m}\right]$.
The index $k$ gives the number of nuclear spins within radius $r_{k}$
of the origin and $N$ is the number of nuclear spins within radius
$l_{0}$, so in $d$ dimensions $\left(\frac{r_{k}}{l_{0}}\right)^{d}=\frac{k}{N}$.
From the definition $A_{k}\propto\left|\psi(r_{k})\right|^{2}$ this
gives the coupling constants $A_{k}=2\exp\left[-\left(\frac{k}{N}\right)^{\frac{m}{d}}\right]$
in units where $A_{0}/2=1$. For times $t\lesssim N^{3/2}/A$, it
is strictly valid to take $N_{\mathrm{tot}}\to\infty$ and change
the sum to an integral in (\ref{eq:IPlusMinusLaplaceTrans}),\cite{coish:2004a}
which gives the final result for $I_{\pm}(s)$, above. 
\begin{figure}
\begin{center}

\includegraphics[%
  bb=130bp 380bp 400bp 730bp,
  clip,
  scale=0.4]{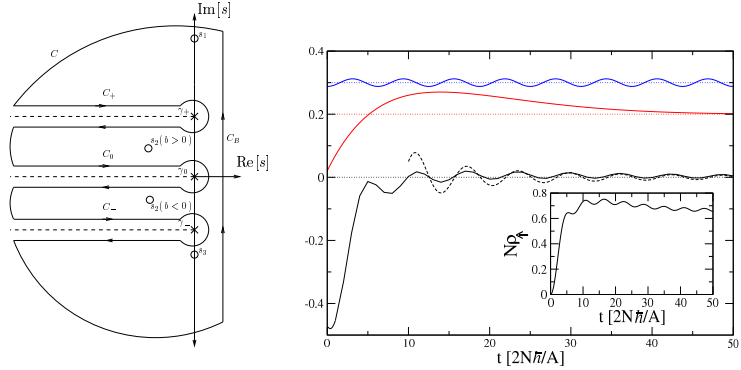}~~~\includegraphics[%
  clip,
  scale=0.25]{fig1b.eps}

\end{center}

\caption{\label{cap:ContourAndDynamics}Contour integral (left) that determines
all contributions (right) to the spin expectation values, including
undamped oscillations (right, top), exponential decay (right, middle)
and nonexponential decay (right, bottom; solid line: numerical integration,
dashed line: analytic asymptotic expression). All three contributions
are added to obtain $\rho_{\uparrow}(t)=1/2+\left\langle S_{z}\right\rangle _{t}$
(right, inset). For these calculations we have chosen $d=m=2$, $I=1/2$,
the initial condition $\rho_{\uparrow}(0)=0$, and values of $b$
and $p$ that correspond to $\Delta=10/11$.}
\end{figure}

The time-dependent spin expectation values can now be recovered from
the Laplace transform expressions by evaluating the Bromwich inversion
integral:\begin{equation}
\left\langle S_{z}\right\rangle _{t}=\frac{1}{2\pi i}\int_{C_{B}}dse^{st}S_{z}(s);\,\,\,\left\langle S_{+}\right\rangle _{t}=\frac{1}{2\pi i}\int_{C_{B}}dse^{st}S_{+}(s).\label{eq:BromwichContour}\end{equation}
 To evaluate these integrals we close the contour in the negative
real half-plane, avoiding all branch cuts of the functions $S_{z}(s),\, S_{+}(s)$,
and apply the residue theorem. The relevant contour is shown in Figure 1 for $S_{+}(s)$ or $S_{z}(s-i\omega_{n})$
within Born approximation when $d=m=2$ (this applies to a two-dimensional
quantum dot with parabolic confinement). The spin expectation values
that result from this procedure are the sum of contributions with
undamped oscillations (from poles on the imaginary axis), exponential
decay (from poles with finite negative real part) and nonexponential
long-time tails (from branch cut integrals). Since the contributions
from poles on the imaginary axis do not decay, the spin expectation
values (in a suitable co-rotating frame) are given generically in
terms of a constant piece, and a remainder with nontrivial dynamics:\begin{equation}
\left\langle S_{X}\right\rangle _{t}=\left\langle S_{X}\right\rangle _{0}+R_{X}(t),\,\,\, X=+,z.\label{eq:SXConstRemainder}\end{equation}
 We show the different contributions to $R_{z}(t)$ in Figure 1
for $d=m=2$ in the weakly perturbative regime (where $\Delta\lesssim1$). 

In the strongly perturbative regime (when $\Delta\ll1$), the time
dependence of $R_{X}(t)$ is given exclusively in terms of the functions
$I_{\pm}(t)=\frac{1}{2\pi i}\int_{C_{B}}dse^{st}I_{\pm}(s)$ that
appear in (\ref{eq:IPlusMinusLaplaceTrans}), above. There is an abrupt
crossover in the long-time behavior of these functions at a critical
value of $d/m$. For $d/m<2$, the major contributions to $I_{\pm}(t\gg N/A)$
come from the upper limit of the integral in (\ref{eq:IPlusMinusLaplaceTrans}),
corresponding to nuclear spins near the center of the quantum dot.
This leads to a modulation of the spin dynamics at a frequency $A/2N$.
When $d/m\ge2$, the major contributions come from the lower limit,
corresponding to nuclear spins far from the dot center. In this case,
the long-time dynamics are smoothly varying, with no modulation: 
\begin{equation}
R_{X}(t)\sim I_{\pm}(t\gg N/A)\propto\left\{ \begin{array}{c}
\left(\frac{1}{t}\right)^{d/m}e^{\pm i\frac{A}{2N}t},\,\,\nu=\frac{d}{m}-1<1\\
\frac{\ln^{\nu}t}{t^{2}},\,\,\nu=\frac{d}{m}-1\ge1\end{array}\right..\label{eq:RxCrossover}\end{equation}
\begin{figure}
\begin{center}
\includegraphics[%
  scale=0.3]{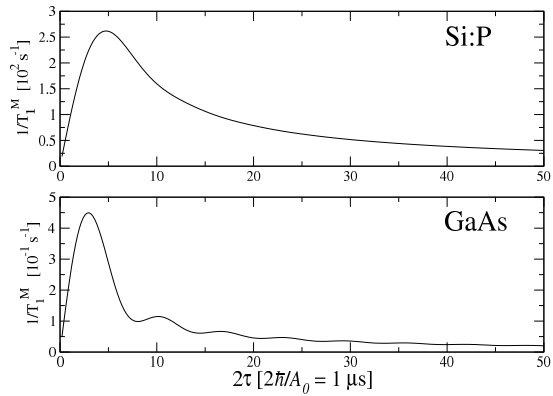}

\end{center}

\caption{\label{cap:CPMGDecayRate}Longitudinal-spin decay rate of the CPMG
echo envelope $(1/T_{1}^{M}\propto R_{z}(2\tau)/2\tau)$ as a function
of the free evolution time $2\tau$ between $\pi$-pulses. We plot
the results for an electron bound to a phosphorus donor, where $N=10^{2}$
(top) and a two-dimensional GaAs quantum dot with $N=10^{5}$ (bottom). }
\end{figure}
The difference in these two cases should be visible in a spin-echo
experiment that uses a Carr-Purcell-Meiboom-Gill (CPMG) spin-echo
sequence: $\frac{\pi}{2}-(\tau-\pi_{x}-\tau-\mathrm{ECHO}-\tau-\pi_{-x}-\tau-\mathrm{ECHO})_{\mathrm{repeat}}$.
We consider the strongly perturbative limit ($\Delta\ll1$), and to
resolve the relevant dynamics, the time between $\pi$-pulses must
satisfy $\tau\ll\sqrt{\delta}\tau_{\mathrm{dd}}$, where $\delta=\Delta^{2}/N$,
and $\tau_{\mathrm{dd}}$ is the nuclear spin dipolar correlation
time.\cite{coish:2004a} Under these conditions the CPMG echo envelope
decay rate as a function of $\tau$ is determined exclusively by the
remainder term according to $1/T_{1}^{M}\propto R_{z}(2\tau)/2\tau$
for the longitudinal component and $1/T_{2}^{M}\propto\mathrm{Re}[R_{+}(2\tau)]/2\tau$
for the transverse components. We plot the CPMG decay rate as a function
of $2\tau$ in Figure \ref{cap:CPMGDecayRate} for two systems of
interest. For an electron trapped at a donor impurity in bulk silicon,
$d=3$ and the orbital wave function is exponential ($m=1$). This
corresponds to $\nu=2$ in (\ref{eq:RxCrossover}). In a two-dimensional
GaAs quantum dot ($d=2$) with parabolic confinement, the ground-state
orbital electron wave function is a Gaussian ($m=2$), which corresponds
to $\nu=0$ in (\ref{eq:RxCrossover}).

\section{Conclusions}

We have reviewed our theoretical description for the dynamics of a
localized electron spin interacting with a nuclear spin environment.
We have predicted a sharp crossover in the relevant dynamics at a
critical value of the dimensionality and form of the electron envelope
wave function, and have described a standard method that could be
used to reveal the relevant dynamics. We stress that the electron
spin dynamics are in general very rich, described by contributions
with exponential decay, nonexponential decay and undamped oscillations.
Furthermore, this work may have profound implications for the future
of spin-based solid-state quantum information processing and quantum
error correction, where previous studies have assumed exponential
decay to zero.

\begin{acknowledgements}
We thank B. L. Altshuler, O. Chalaev, H.-A. Engel, S. Erlingsson,
H. Gassmann, V. Golovach, A. V. Khaetskii, F. Meier, D. S. Saraga,
J. Schliemann, N. Shenvi, L. Vandersypen, and E. A. Yuzbashyan for
useful discussions. We acknowledge financial support from the Swiss
NSF, the NCCR nanoscience, EU RTN Spintronics, DARPA, ARO, ONR and
NSERC of Canada.\vspace{-0.5cm}
\end{acknowledgements}
\bibliographystyle{/usr/share/texmf/bibtex/bst/kluwer/klunum}
\bibliography{/home/coish/papers/hfi2004/hfBibliography}
\end{article}
\end{document}